%Paper: cond-mat/9404054
%From: Antonio Murilo <amsm@thphys.ox.ac.uk>
%Date: Mon, 18 Apr 94 15:38:03 BST
%Date (revised): Mon, 18 Apr 94 17:02:50 BST

\magnification=\magstep1
\hsize=128mm
\baselineskip=18pt
\vglue 1in

\centerline{\bf INTENSITY CORRELATIONS IN ELECTRONIC WAVE PROPAGATION IN A}
\centerline{\bf  DISORDERED MEDIUM: THE INFLUENCE OF SPIN-ORBIT SCATTERING}
\vskip 10mm

\centerline{A. M. S. Mac\^edo}

\centerline{\it Theoretical Physics, University of Oxford,}

\centerline{\it 1 Keble Road, Oxford OX1 3NP, U.K.}
\vskip 10mm

\centerline{\bf Abstract}
\medskip
We obtain explicit expressions for the correlation functions of transmission
and reflection coefficients of coherent electronic waves propagating through a
disordered quasi-one-dimensional medium with purely elastic diffusive
scattering in the presence of spin-orbit interactions. We find in the metallic
regime both large local intensity fluctuations and long-range correlations
which ultimately lead to universal conductance fluctuations. We show that the
main effect of spin-orbit scattering is to suppress both local and long-range
intensity fluctuations by a universal symmetry factor 4. We use a scattering
approach based on random transfer matrices.
\medskip

\vfill\eject
\noindent
{\bf I. Introduction}
\medskip
There has been considerable effort made in the past few years to understand the
statistical properties of the intensity pattern created by coherent waves
propagating through a disordered medium in the multiple elastic scattering
regime. The simplest known effect is large local intensity fluctuations (the
speckle pattern) observed in scattering of light [1], which can be interpreted
theoretically [2] as being due to fluctuations in the sum of a large number of
uncorrelated amplitudes associated with different Feynman paths arising from
the diffusive motion of the scattered wave.

Another interesting feature of such patterns, which has also been
experimentally observed [3,4], is the presence of coherent backscattering. A
result interpreted as a constructive interference between pairs of
time-reversed scattering paths.

It is now understood that the statistical properties of the fluctuating
intensity patterns may crucially depend on the complex interference of the
underlying wave fields of the diffusion modes in the bulk of the medium. This
dependence leads to non-trivial long-range correlations [5,6], which in turn
imply fluctuations of the total transmission coefficient of order unity. Such
fluctuations have been observed in phase coherent electronic waves transmitted
across a disordered  sample [7-10]. The theory of this effect [11-15] predicts
that the conductance of a disordered mesoscopic metallic conductor fluctuates
with an amplitude always of order $e^2/h$ independent of the system size and
degree of disorder, the phenomenon of universal conductance fluctuations.

An important consequence of this result is that it implies strong correlations
between diffusion modes and thus reflected and transmited intensities in
different modes must be statistically correlated.

All these interesting statistical properties are, in fact, expected to occur in
any system in which waves propagate coherently in a disordered medium.
Electronic waves, though, can give rise to special quantum mechanical effects
if the impurity scattering becomes spin-dependent by adding to the sample
strong spin-orbit couplers like Au. The simplest observable effect is the
depletion of backscattering [16], which results from coherent destructive
interference of spin-dependent wave functions associated with pairs of
time-reversed scattering events. Theoretically, it has been shown [17] that
spin-orbit scattering supresses universal conductance fluctuations by a factor
of 4 and gives rise to weak anti-localization, slightly enhancing the value of
the average conductance.

There are two different approaches to coherent electronic wave propagation in
disordered medium. The first one [11-13] (the microscopic approach) is based on
linear response theory and uses Feynman diagrams to perform averages over
different impurity configurations. The second one [15,18] (the macroscopic
approach) is based on Landauer scattering theory [19] and uses random transfer
matrices to represent the multiple elastic scattering of electronic waves in
the sample. These two approaches are essentially equivalent for systems with
quasi-one-dimensional geometry.

Explicit expressions for the correlation functions of the transmission
coefficients of waves propagating through a random medium with purely elastic
scattering have recently been obtained by Feng {\it et al.} [6] using the
microscopic approach. They obtained, in addition to the familiar speckle
pattern, long-range correlations resulting from the interaction of diffusion
modes. A similar result has also been derived by Mello {\it et al.} [20] using
the macroscopic approach. They showed that long-range correlations between
reflection coefficients are of the same order as that between transmission
coefficients, a result which had some importance in connection with a deeper
understanding of universal conductance fluctuations.

This paper is concerned with correlations of transmission and reflection
coefficients of electronic waves diffusively scattered in a disordered medium
both with and without strong spin-orbit coupling. To the best of our knowledge
there is no explicit evaluation of correlation functions of transmission and
reflection coefficients in the presence of spin-orbit scattering. In this paper
we provide such a calculation together with results for a correlation function
not considered in previous works.
In presence of spin-dependent scattering a number of new purely quantum
mechanical interference effects takes place and the above
 authors results do not apply, even locally, where we found the amplitude of
intensity fluctuations to change dramatically.  We use the macroscopic approach
to obtain results for the unitary, orthogonal and symplectic ensembles. The
unitary ensemble is appropriate to systems with an applied magnetic field. The
orthogonal ensemble describes systems with time-reversal symmetry in the
absence of spin-orbit scattering, whilst the symplectic ensemble applies to
systems with time-reversal symmetry in the presence of strong spin-orbit
scattering.
In Sec. II we formulate the scattering problem and show how the various
quantities of interest can be written in terms of some parameters that appear
in the polar decomposition of the transfer matrix $M$ of the disordered sample.
We assume the system to be quasi-one-dimensional, which greatly simplifies the
calculations owing to the isotropic nature of the probability distribution
$P(M)$ in the group manifold of the transfer matrix.
In Sec. III we obtain novel exact expressions for the various correlation
functions of reflection and transmission coefficients in terms of the number
$N$ of transverse propagating modes and certain averages to be performed by
solving an isotropic diffusion equation for $P(M)$, in which the length $L$ of
the sample plays the role of an embedding parameter. In the metallic regime we
find that large local fluctuations and long-range correlations are suppressed
by coherent interferences generated by spin-orbit scattering. We show that our
result for the speckle pattern can be  understood in terms of a sum of a large
number of uncorrelated Feynman paths. The importance of long-range correlations
to universal conductance fluctuations is briefly discussed. We present a
summary and conclusions in Sec. IV.
\bigskip
\noindent
{\bf II. The Scattering Approach}
\medskip
We consider a confined quasi-one-dimensional geometry in which semi-infinite
perfectly conducting leads are attached to the right and left sides of a
disordered region of length $L$ and width $W$. The total number of transverse
propagation modes is given by $N \sim (k_F W)^{d-1}$, where $k_F$ is the Fermi
momentum and $d$ is the sample dimension. We assume $L \ll L_{in}$, where
$L_{in}$ is the inelastic scattering length, thus the only important scattering
mechanism is elastic and the electron wave function is coherent.

Since the orthogonal and unitary ensembles have been thoroughly discussed in
many recent papers [15,18], we present a brief review only of the less familiar
symplectic case.

The amplitude of the propagating modes to the left of the sample are related to
those to the right by the transfer matrix $M$. Flux conservation and
time-reversal symmetry in the presence of spin-orbit scattering imply that $M$
can conveniently be parametrized as [17]
$$
M=\pmatrix{u  &  0   \cr
           0  &  u^* \cr}
\pmatrix{\sqrt{1+\lambda}  &  \sqrt{\lambda}   \cr
         \sqrt{\lambda}    &  \sqrt{1+\lambda} \cr}
\pmatrix{v  &  0   \cr
         0  &  v^* \cr},
\eqno (2.1)
$$
where  $u$ and $v$ are arbitrary $N\times N$ quaternion unitary matrices and
$\lambda$ is an $N\times N$ quaternion real, diagonal matrix whose elements
$\lambda_1, \lambda_2,\dots, \lambda_N$ define the eigenchannels.
The operations over quaternion matrices and elements are defined in Ref. 17.

The transfer matrix $M$ can also be written in terms of transmission and
reflection matrices $t,r$ (fluxes incident from the left) and $t',r'$ (fluxes
incident from the right). It reads
$$
M=\pmatrix{(t^{\dagger})^{-1} & r'(t')^{-1} \cr
           -(t')^{-1}r & (t')^{-1} \cr}.
\eqno (2.2)
$$
Comparing (2.2) with (2.1) one finds
$$
r=-\bar v\left({\lambda \over 1+\lambda}\right)^{{\scriptstyle 1\over
\scriptstyle 2}} v,
\eqno (2.3)
$$
$$
t= u\left({1 \over 1+\lambda}\right)^{{\scriptstyle 1\over \scriptstyle 2}} v.
\eqno (2.4)
$$
The reflection and transmission coefficients $R_{ab}$ and $T_{ab}$ of wave
amplitudes in channel $a$ when there is an incident flux in channel $b$ are
given by [21]
$$
R_{ab}^{}=\left|r_{ab}^{}\right|^2 \equiv \sum_{i=0}^3 |
r_{ab}^{(i)}|^2,
\eqno (2.5)
$$
$$
T_{ab}^{}=\left|t_{ab}^{}\right|^2 \equiv \sum_{i=0}^3 |
t_{ab}^{(i)}|^2,
\eqno (2.6)
$$
in which $r_{ab}^{(i)}$ and $t_{ab}^{(i)}$ are complex components of the
quaternions $r_{ab}$ and $t_{ab}$ in the standard basis $\{ e_{i}, i=0,1,2,3
\}$, where $e_0$ is the $2\times 2$ unit matrix and $e_1, e_2, e_3$ are defined
as
$$
e_1=\pmatrix{i &  0 \cr
             0 & -i \cr},
\quad
e_2=\pmatrix{0 & -1 \cr
             1 &  0 \cr},
\quad
e_3=\pmatrix{0  &  -i \cr
             -i &  0 \cr}.
\eqno (2.7)
$$
We define three kinds of crossed second moment
$$
C_{ab,a'b'} \equiv \langle T_{ab} T_{a'b'} \rangle -
\langle T_{ab} \rangle \langle T_{a'b'} \rangle,
\eqno (2.8)
$$
$$
D_{ab,a'b'} \equiv \langle R_{ab} R_{a'b'} \rangle -
\langle R_{ab} \rangle \langle R_{a'b'} \rangle,
\eqno (2.9)
$$
$$
E_{ab,a'b'} \equiv \langle R_{ab} T_{a'b'} \rangle -
\langle R_{ab} \rangle \langle T_{a'b'} \rangle,
\eqno (2.10)
$$
in which $\langle \ldots \rangle$ stands for an average over an ensemble of
random matrices $M$ of the form (2.1). It has been shown [17] that, as the
sample length is increased, $M$ performs an isotropic random walk in its group
manifold. This implies that ensemble averages of arbitrary functions of the
form $f(M)=f_1(u,v)f_2(\lambda)$ are simply given by
$$
\langle f(M) \rangle = \langle f_1(u,v) \rangle_0^{}
\langle f_2(\lambda) \rangle_s^{},
\eqno (2.11)
$$
where ${\langle f_1 \rangle}_{\smash 0}^{\smash{}}$ denotes the average on the
quaternion unitary group and ${\langle  f_2 \rangle}_{\smash s}^{\smash{}}$ is
the expectation value obtained by solving the evolution equation
$$
{\partial
\langle f_2\rangle_{\smash s}^{\smash{}}
 \over \partial s}
={1 \over 2N-1}
\Biggl\langle {1 \over J}
 \sum_{a=1}^N {\partial \over \partial \lambda_a}
\biggl[\lambda_a (1+\lambda_a)J
{\partial f_2 \over \partial \lambda_a} \biggr]
\Biggr\rangle_{ \displaystyle s}^{\smash{}} ,
\eqno (2.12)
$$
in which $J=\prod_{a<b}^{}\left|{\lambda_a-\lambda_b}\right|^{4}$ and $s$ is
the length of the sample in units of the mean free path.

The quasi-one-dimensional geometry owes its mathematical simplicity to the
factorization shown in (2.11).
\bigskip
\noindent
{\bf III. The Correlation Functions}
\medskip
If we insert (2.6) into (2.8), use (2.11) and perform the averages on the
quater-
\eject
\noindent
nion unitary group in the way described in Refs. 17 and using the results of
Ref. 22, we find
$$\eqalignno{
C_{ab,a'b'}=&(A_{C,1}\langle T^2 \rangle^{}_{s}-A_{C,2}\langle T_2
 \rangle^{}_{s})(4+ \delta_{aa'}\delta_{bb'})\cr
&+(A_{C,1}\langle T_2 \rangle^{}_{s}-4A_{C,2}\langle T^2
 \rangle^{}_{s})(\delta_{aa'}+\delta_{bb'})
-\langle T_{ab} \rangle \langle T_{a'b'} \rangle ,
&(3.1) \cr
}
$$
where
$$
A_{C,1}={4N^2+1 \over N^2 (4N^2-1)^2},
\qquad
A_{C,2}={2 \over N(4N^2-1)^2},
\eqno (3.2)
$$
$$
T=\sum_{ab} T_{ab}=\sum_{a=1}^N {1 \over 1+\lambda_a},
\qquad
T_2=\sum_{a=1}^N {1 \over (1+\lambda_a)^2}.
\eqno (3.3)
$$
Equation (3.1) is exact. As a check one can verify that
$$
{\rm Var}(T)=\sum_{aba'b'}C_{ab,a'b'}
\eqno (3.4)
$$
holds for arbitrary $N$.

Inserting (2.5) into (2.9) and performing the averages on the quaternion
unitary group we get
$$\eqalignno{
D_{ab,a'b'}=&A_{D,1}(\langle R_2 \rangle^{}_{s} -\langle R^2 \rangle^{}_{s})
\delta_{aba'b'}+(A_{D,2}\langle R^2 \rangle^{}_{s}-A_{D,3}\langle R_2
\rangle^{}_{s})((2-\delta_{ab})(2-\delta_{a'b'}) \cr
&+\delta_{aa'}\delta_{bb'}+\delta_{ab'}\delta_{ba'})+
2(A_{D,3}\langle R^2 \rangle^{}_{s}-A_{D,4}\langle R_2 \rangle^{}_{s})
(\delta_{aa'b'}+\delta_{abb'}+\delta_{aba'}\cr
&+\delta_{ba'b'}-\delta_{aa'}-\delta_{ab'}-\delta_{bb'}-\delta_{ba'})
-\langle R_{ab} \rangle \langle R_{a'b'} \rangle ,
&(3.5) \cr
}
$$
in which
$$
A_{D,1}={1 \over N(N-1)(2N-3)(2N-1)},
\eqno (3.6a)
$$
$$
A_{D,2}={2N^2-3N+2 \over N^2(N-1)(2N-3)(4N^2-1)},
\eqno (3.6b)
$$
$$
A_{D,3}={1 \over N^2(N-1)(2N-3)(2N+1)},
\eqno (3.6c)
$$
$$
A_{D,4}={2N^2-N+1 \over 2N^2(N-1)(2N-3)(4N^2-1)},
\eqno (3.6d)
$$
and
$$
R=\sum_{ab} R_{ab}=\sum_{a=1}^N {\lambda_a \over 1+\lambda_a},
\qquad
R_2=\sum_{a=1}^N \left({\lambda_a \over 1+\lambda_a}\right)^2.
\eqno (3.7)
$$
The unit tensors $\delta_{abc}$ and $\delta_{abcd}$ take the value 1 if all
indices coincide and vanish otherwise. One can easily check that $D_{ab,a'b'}$
satisfies
$$
{\rm Var}(R)=\sum_{aba'b'}D_{ab,a'b'}.
\eqno (3.8)
$$
Finally for (2.10) one finds
$$\eqalignno{
E_{ab,a'b'}=&(A_{E,1}\langle R T \rangle^{}_{s} -
A_{E,2}(\langle T \rangle^{}_{s}-\langle T_2 \rangle^{}_{s}))
(2-\delta_{ab})
+(A_{E,3}(\langle T \rangle^{}_{s}-\langle T_2 \rangle^{}_{s})
 \cr
&
-A_{E,2}\langle R T \rangle^{}_{s})(\delta_{ab'}+\delta_{bb'}
-\delta_{abb'})
-\langle R_{ab} \rangle \langle T_{a'b'} \rangle ,
&(3.9) \cr
}
$$
where
$$
A_{E,1}={1 \over N^2(2N+1)(N-1)},
\qquad
N A_{E,2}=A_{E,3}={1 \over N(N-1)(4N^2-1)}.
\eqno (3.10)
$$
The averages $\langle R_{ab} \rangle$ and $\langle T_{ab} \rangle$ have been
calculated in Ref. 17 and are given by
$$
\langle R_{ab} \rangle={2-\delta_{ab} \over N(2N-1)}\langle R \rangle^{}_{s},
\qquad
\langle T_{ab} \rangle={\langle T \rangle^{}_{s}\over N^2}.
\eqno (3.11)
$$
Observe that $\langle R_{aa} \rangle \le \langle R_{ab} \rangle$ in agreement
with weak-localization theory [16]. One can verify that $E_{ab,a'b'}$ satisfies
$$
{\rm Cov}(R,T)=\sum_{aba'b'}E_{ab,a'b'}.
\eqno (3.12)
$$
Equations (3.1), (3.5) and (3.9) are new predictions of the present macroscopic
model. Comparing equations (3.1) and (3.5) with the ones obtained in Refs.
15 and 20 for the unitary and orthogonal ensembles we find great similarities
in their structure. We remark that the correlation function $E_{ab,a'b'}$ has
not been considered in earlier works.

To proceed further we have to evaluate the remaining averages using the
evolution equation (2.12). Since no exact solution is possible for arbitrary
$s$, we confine ourselves to the important metallic or diffusive regime,
defined as the one in which $1 \ll s \ll N$. In this regime the average total
transmission $\langle T \rangle^{}_s$, which is proportional to the
conductance, has a very large value and it makes sense to expand an arbitrary
average $\langle f(\lambda) \rangle^{}_s$ in inverse powers of $\langle T
\rangle^{}_s$. This procedure is explained thoroughly in Ref 15. We merely
quote here the resulting expressions for $C_{ab,a'b'}$, $D_{ab,a'b'}$ and
$E_{ab,a'b'}$, which are
$$
C_{ab,a'b'}=\langle T_{ab} \rangle \langle T_{a'b'} \rangle
\left({1 \over \alpha} \delta_{aa'}\delta_{bb'}+{2 \over 3\alpha\langle T
\rangle^{}_s}(\delta_{aa'}+\delta_{bb'}) +
{2 \over 15 \beta \langle T \rangle^{2}_s} \right),
\eqno (3.13)
$$
$$\eqalignno{
D_{ab,a'b'}= & ~\langle R_{ab} \rangle \langle R_{a'b'} \rangle
\biggl(
{\beta \over 2\alpha}
\left(
{ 2\delta_{aa'}\delta_{bb'}+\gamma\delta_{ab'}\delta_{ba'} \over
\beta + \delta_{ab}(2-\beta) }
\right) \cr
&~~~~~~~~~~\cr
&-
{(2(\delta_{aa'}+\delta_{bb'})+\gamma(\delta_{ab'}+\delta_{ba'}
-\beta \delta_{aa'bb'}))\over 2\alpha \langle R \rangle^{}_s }
+{32 \over 15 \beta \langle R \rangle^{2}_s}
\biggr),~~~~~~
&(3.14) \cr
}
$$
$$
E_{ab,a'b'}=-\langle R_{ab} \rangle \langle T_{a'b'} \rangle
\left(
{ 2\delta_{bb'}+\gamma\delta_{ab'} \over
3\alpha\langle R \rangle^{}_s }
+{2 \over 15 \beta \langle R \rangle^{}_s \langle T \rangle^{}_s}
\right),~~~~~~~~~~~~~~~~
\eqno (3.15)
$$
where $\alpha=(\beta^2-3\beta+4)/2$, $\gamma=(2-\beta)(3-\beta)$ and $\beta$
takes on the value 4 in the symplectic ensemble. Other possible values are
$\beta=1$ (orthogonal ensemble) and $\beta=2$ (unitary ensemble), which have
been studied in Refs. 15 and 20. Expression (3.15) is, however, for all values
of $\beta$, a new result of the present work.

Suppose a single mode $b$ is excited on the left and we are interested in
correlations between transmission coefficients in different modes $a$ and $a'$
on the right. From (3.13) we get
$$
{\rm Cov}(T_{ab},T_{a'b})={1 \over \alpha}\langle T_{ab}\rangle
\langle T_{a'b} \rangle \delta_{aa'},
\eqno (3.16)
$$
thus the present model predicts large local fluctuations of transmission
intensities. For conventional waves, $\alpha=1$, and (3.16) is the familiar
speckle pattern observed in many experiments [1]. For electronic waves in the
presence of spin-orbit scattering, $\alpha=\beta=4$, and from (3.16) we see
that local fluctuations are suppressed by a factor of 4.

The above result can be easily understood by means of a very simple argument.
The presence of disorder implies that the electronic motion is diffusive and
thus the transmission amplitude $t_{ab}$ is a large sum of amplitudes
associated with Feynman paths that cover the whole sample. So, we may write
$$
t_{ab}=\sum_{i=1}^M A_{ab}(i),
\eqno (3.17)
$$
in which $A_{ab}(i)$ is the probability amplitude due to the i-th path
connecting the channel $a$ to the channel $b$. We assume that $A_{ab}(i)$ are
independent random quaternion variables uniformly distributed on the group
$U(2)$. Inserting (3.17) into (2.6) we find
$$
\langle T_{ab}\rangle=\sum_{i} \langle | A_{ab}(i) |^2 \rangle,
\eqno (3.18)
$$
while
$$\eqalignno{
\langle T_{ab}^2\rangle=&{5 \over 4}\sum_{ij} \langle | A_{ab}(i) |^2
| A_{ab}(j) |^2\rangle \cr
\simeq &{5 \over 4} \langle T_{ab}\rangle ^2,
&(3.19) \cr
}
$$
where terms of order unit compared with $M$ were neglected. Using (3.19) we
easily find
$$
{\rm Var}(T_{ab})={1 \over 4} \langle T_{ab} \rangle^2,
\eqno (3.20)
$$
in agreement with (3.16).

The second term in (3.13) is a long-range correlation which becomes dominant
when we consider the fluctuations of the total transmission coefficient
$T_b\equiv \sum_a T_{ab}$ to the right due to a single mode $b$ excited on the
left. From (3.13) we get
$$
{\rm Var}(T_b)={2 \over 3\alpha N s}.
\eqno (3.21)
$$
When $\alpha=1$ (conventional waves), (3.21) is the same result obtained in
Refs. 5,6 and 19. In the symplectic case ($\alpha=4$), the present model
predicts suppression of long-range correlations by the same factor 4 found in
(3.16).

The third term in (3.13) describes a uniform long-range correlation between all
channels. It is the dominant fluctuation of the total transmission coefficient
$T$, which is related to the conductance by the Landauer formula [19] $G=2T$.
{}From (3.13) we find
$$
{\rm Var}(T)={2 \over 15 \beta},
\eqno (3.22)
$$
in agreement with microscopic calculations [11-15].

This result shows that the long-range nature of the diffusive process induces
non-trivial correlations between transmission coefficients, which in turn
generates universal fluctuations as observed in the conductance of phase
coherent conductors [7-10].

Let us now consider (3.14). The first term gives for $\beta=1$ and $\beta=2$
$$
{\rm Var}(R_{ab})=\langle R_{ab} \rangle^2,
\eqno (3.23)
$$
which is the familiar speckle pattern for the reflection coefficient of
conventional waves. In the symplectic case ($\beta=4$), we get
$$
{\rm Var}(R_{ab})={1 \over 4}\langle R_{ab} \rangle^2,
\qquad {\rm for}~~~ a \ne b,
\eqno (3.24)
$$
and
$$
{\rm Var}(R_{aa})=\langle R_{aa} \rangle^2,
\eqno (3.25)
$$
Equation (3.24) can be easily derived following the same reasoning used to
obtain (3.20). We shall consider the more subtle result shown in (3.25).

The Feynman paths that contribute to $r_{aa}$ are closed, which implies that
interference between time reversed trajectories plays an important role and
gives rise to coherent backscattering. We include this information in $r_{aa}$
by writting it as
$$
r_{aa}=\sum_{i=1}^M A_{aa}(i) \bar A_{aa}(i),
\eqno (3.26)
$$
in which $A_{aa}(i)$ is the probability amplitude associated with the i-th
Feynman  path that starts and finishes at channel $a$, whilst $\bar A_{aa}(i)$
is the probability amplitude due to the time-reversed path. We assume again
that $A_{aa}(i)$ are random and independent quaternion variables and that the
number of paths $M$ is large. Inserting (3.26) into (2.5) we find
$$
\langle R_{aa}\rangle=\sum_{i} \langle | A_{aa}(i) |^4 \rangle,
\eqno (3.27)
$$
while
$$\eqalignno{
\langle R_{aa}^2\rangle=&~2\sum_{ij} \langle | A_{aa}(i) |^4| A_{aa}(j) |^4
\rangle \cr  \simeq &~2 \langle R_{aa}\rangle^2,
& (3.28) \cr
}
$$
which of course implies (3.25).

It is very instructive to observe that spin-orbit effects on local
fluctuations, whilst being intrinsically of quantum mechanical nature, can be
understood following the same kind of reasoning that explains the familiar
speckle pattern of conventional waves.

Another noteworthy feature of (3.14) is that it shows long-range
correlations which are of the same order of magnitude as those in (3.13). This
fact was discussed in Ref. 20 to contrast with earlier hypothesis [2], which
suggested that absence of long-range correlations between reflection
coefficients might be consistent with universal conductance fluctuations. One
can see from (3.14) that it is precisely these long-range correlations that
give rise to universal fluctuations. We find
$$
{\rm Var}(R)={2 \over 15 \beta},
\eqno (3.29)
$$
which is the same value found in (3.22). This is consistent with the identity
$R=N-T$.

Consider now Eq. (3.15). Observe that it shows only long-range correlations,
since reflected and transmitted waves emerge at different sides of the sample.
Such correlations exist simply because  the interference pattern generated by
the scattered wave is highly sensitive to any small change in the impurity
configuration [23].

When a single channel $b$ is excited on the left, we find that the total
transmission coefficient to the right $T_b$ is correlated with the total
reflection coefficient to the left $R_b$ with covariance given by
$$
{\rm Cov}(T_b,R_b)=-{2 \over 3\alpha N s}.
\eqno (3.30)
$$
Comparing (3.30) with (3.21) we find
$$
{\rm Cov}(T_b,R_b)=-{\rm Var}(T_b),
\eqno (3.31)
$$
which is consistent with the identity $R_b+T_b=1$.

Finally, the second term in (3.15) is a uniform background correlation that
gives rise to a universal correlation between the total transmission
coefficient $T$ and the total reflection coefficient $R$. Using (3.12)and
(3.15) we find
$$
{\rm Cov}(R,T)=-{2 \over 15 \beta},
\eqno (3.32)
$$
which is consistent with (3.22) and the identity $R+T=N$.
\bigskip
\noindent
{\bf IV. Summary and Conclusions}
\medskip
In this work we have studied the correlations between transmission and
reflection coefficients of coherent electronic waves diffusively scattered in a
disordered quasi-one-dimensional medium both with and without spin-orbit
scattering. In the absence of spin-orbit scattering the quantum mechanical
nature of the phenomena enters only via the wave-like behavior of the electron
and thus the same statistical properties of transmitted and reflected
intensities are expected to occur in any system in which waves propagate
coherently through a disordered medium. In the presence of spin-orbit
scattering, however, new purely quantum mechanical effects take place and
special statistical properties arise.

We have obtained novel exact expressions for several correlation functions
between transmission and reflection coefficients in terms of the number $N$ of
eigenchannels and certain averages to be evaluated by solving a given evolution
equation, in which the length $L$ of the sample plays the role of the diffusion
time.

In the metallic regime, which is defined as the one in which the length of the
sample is much larger than the mean free path, but much smaller than the
localization length, we have found explicit expressions for the various
correlation functions between the reflection and transmission coefficients
valid for the unitary, orthogonal and symplectic ensembles. We have shown that
in the symplectic ensemble special quantum interference effects due to
spin-orbit scattering suppresses both local fluctuations and long-range
correlations of transmission and reflection coefficients.

The high sensitivity of the interference pattern created by the scattered wave
to any small change in the impurity configuration leads to interesting
correlations between the total reflection coefficient
 and the total transmission coefficient, evaluated at opposite  sides of the
sample, when a single wave mode is excited. We believe that this effect is
unique to the diffusive regime and plays an important role in providing an
understanding of universal fluctuations of global quantities.

We conclude by remarking that the metallic or diffusive regime is very special
in that the interferences of the wave fields of the diffusion modes gives rise
to remarkable statistical properties of transmission and reflection
coefficients some of which are peculiar to the symplectic ensemble. All
long-range correlations that we have found in crossed second moments of
transmission and reflection coefficients have the same order of magnitude and
are ultimately responsable for universal fluctuations of global quantities like
total reflection and total transmission coefficients. The latter corresponds to
the familiar universal conductance fluctuations observed in phase coherent
metallic samples.
\bigskip
\noindent
{\bf Acknowledgments}
\medskip
The author is grateful to J. T. Chalker for helpful comments and discussions.
This work was partially supported by CAPES (Brazilian Agency).
\bigskip
\noindent
{\bf References}
\medskip
{\parindent=0.5cm
\item{[1]} S. Etemad, R Thompson and H. J. Andrejico, Phys. Rev. Lett. {\bf
57}, 575 (1986); M. Kaveh, M. Rosenbluh, I. Edrei and I. Freund, Phys. Rev.
Lett. {\bf 57}, 2049 (1986).
\item{[2]} P. A. Lee, Physica (Amsterdam) {\bf 140 A}, 169 (1986).
\item{[3]} M. P. van Albada and A. Lagendijk, Phys. Rev. Lett. {\bf 55}, 2692
(1985).
\item{[4]} P. E. Wolf and G. Maret, Phys Rev. Lett. {\bf 55}, 2696 (1985).
\item{[5]} M. J. Stephen and G. Cwilich, Phys Rev. Lett. {\bf 59}, 285 (1987).
\item{[6]} S. Feng, C. Kane, P. A. Lee and A. D. Stone, Phys Rev. Lett. {\bf
61}, 834 (1988).
\item{[7]} C. P. Umbach, S. Washburn, R. B. Laibowitz, and R. A. Webb, Phys.
Rev. B {\bf 30}, 4048 (1984).
\item{[8]} R. A. Webb, S. Washburn, C. P. Umbach, and R. B. Laibowitz, Phys.
Rev. B {\bf 54}, 2696 (1985).
\item{[9]} S. Washburn and R. A. Webb, Adv. Phys. {\bf 35}, 375 (1986).
\item{[10]} A. G. Aronov and Yu. V. Sharvin, Rev. Mod. Phys. {\bf 59}, 755
(1987).
\item{[11]} P. A. Lee and A. D. Stone, Phys. Rev. Lett. {\bf 55}, 1622 (1985);
B. L. Altshuler, Pis'ma Zh. Eksp. Teor. Fiz. {\bf 41}, 530 (1985)     [JEPT
Lett. {\bf 41}, 648 (1985)].
\item{[12]} P. A. Lee and T. V. Ramakrishnan, Rev. Mod. Phys. {\bf 57}, 287
(1985).
\item{[13]} P. A. Lee, A. D. Stone, and H. Fukuyama, Phys. Rev. B {\bf 35},
1039 (1987).
\item{[14]} S. Feng,  Phys. Rev. B {\bf 39}, 8722 (1989).
\item{[15]} P. A. Mello, Phys. Rev. Lett. {\bf 60}, 1089 (1988); P. A. Mello
and A. D. Stone, Phys. Rev. B {\bf 44}, 3559 (1991).
\item{[16]} G. Bergmann, Phys. Rep. {\bf 107}, 1 (1984).
\item{[17]} A. M. S. Mac\^edo and J. T. Chalker, Phys. Rev. B {\bf 46}, 14985
(1992).
\item{[18]} P. A. Mello, P. Pereyra, and N. Kumar, Ann. Phys. (N.Y.) {\bf 181},
290 (1988).
\item{[19]} R. Landauer, Philos. Mag. {\bf 21}, 863 (1970).
\item{[20]} P. A. Mello, E. Akkermans, and B. Shapiro, Phys. Rev. Lett. {\bf
61}, 459 (1988).
\item{[21]} In Ref. 17 it was found convenient to define $R_{ab}$ and $T_{ab}$
as the quaternions $R_{ab}^{}=r^{}_{ab}r^\dagger_{ab}$ and
$T_{ab}^{}=t^{}_{ab}t^\dagger_{ab}$, since quaternion random phases vanish for
simple averages like $\langle R_{ab}\rangle$ and  $\langle T_{ab}\rangle$. In
order to describe fluctuations and correlations, however, such simple
definitions do not apply and (2.5) and (2.6) must be used instead.
\item{[22]} P. A. Mello, J. Phys. A {\bf 23}, 4061 (1990).
\item{[23]} S. Feng, P. A. Lee and A. D. Stone, Phys. Rev. Lett. {\bf 56}, 1960
(1986).

\bye